\documentclass[conference]{IEEEtran}

\usepackage{ifpdf}
\usepackage{cite}
\usepackage{circledsteps}
\usepackage{verbments}

\ifCLASSINFOpdf
   \usepackage[pdftex]{graphicx}
\else
  \usepackage[dvips]{graphicx}
\fi

\usepackage{stfloats}
\usepackage{url}
\usepackage{caption} 
\usepackage{tabularx}

\begin{document}
\title{FPsPIN: An FPGA-based Open-Hardware Research Platform for Processing in the Network}


 \author{\IEEEauthorblockN{Timo Schneider}
 \IEEEauthorblockA{ETH Zurich\\
 timos@inf.ethz.ch}
 \and
 \IEEEauthorblockN{Pengcheng Xu}
 \IEEEauthorblockA{ETH Zurich\\
 pengcheng.xu@inf.ethz.ch}
 \and
 \IEEEauthorblockN{Torsten Hoefler}
 \IEEEauthorblockA{ETH Zurich\\
htor@inf.ethz.ch}}


\maketitle \vspace{-2cm}
\begin{abstract}
In the era of post-Moore computing, network offload emerges as a solution to two challenges: the imperative for low-latency communication and the push towards hardware specialization.
Various methods have been employed to offload protocol- and data-processing onto network interface cards (NICs), from firmware modification to running full Linux on NICs for application execution. The sPIN project enables users to define handlers executed upon packet arrival. While simulations show sPIN's potential across diverse workloads, a full-system evaluation is lacking. This work presents FPsPIN, a full FPGA-based implementation of sPIN. FPsPIN is showcased through offloaded MPI datatype processing, achieving a 96\% overlap ratio. FPsPIN provides an adaptable  open-source research platform for researchers to conduct end-to-end experiments on smart NICs.
\end{abstract}


\IEEEpeerreviewmaketitle

\section{Introduction}

Current high-performance computing (HPC) systems use remote direct memory access (RDMA)~\cite{rdma} to remove the central processing unit (CPU) from the network data-path. This allows sub-microsecond latency and 800 GBit transfer rate~\cite{datacenter_ethernet}. The data-steering capabilities of RDMA are limited, data is simply deposited into the virtual memory of the receiver. If data-processing with low computational intensity is required, e.g., an MPI (all)reduce operation or MPI derived-datatype (un)pack, high data rates lead to a performance problem: A modern CPU requires approximately 10 ns to access a 64 bit wide L3 cache line~\cite{molka2014main}. However, at 400 Gbit/s the arrival interval for such a data portion can be as low as 1.2 ns. As a result multiple attempts have been made to offload lightweight data-processing and receiver-side data-steering from the CPU into the network interface card (NIC) or into the network itself~\cite{mogul2003tcp, de2021flare, istvan2016consensus}, leading to the emergence of smart NICs (sNIC).

Offloading protocol- and data-processing can be done in different ways, with implications on ease of use and performance. While the earliest smart NICs allowed users to change the NICs firmware, many modern smart NICs run a full operating system on a separate system-on-chip (SoC) within the NIC, allowing the user to program the SoC without interfering with the core NIC functionality.
We refer to Section~\ref{sec:relwork} for an overview of network offload solutions. While all these approaches have been shown to be useful in specific scenarios, none of them have gained market-wide adoption. The programming models and interfaces being offered are vendor- or even hardware-specific, which hinders research progress in the field of smart NICs.

In contrast, the sPIN abstract machine model~\cite{spin_orig} (see Section~\ref{sec:spin_model}) is hardware agnostic. An open-source version of a sPIN packet processor has been developed~\cite{pspin} and evaluated for numerous workloads in simulations~\cite{pspin, pspin_protobuf, pspin_fs, pspin_ddt} in which incoming packets are fed to a verilator based simulator~\cite{snyder2018verilator}.

\begin{figure}[!ht]
\centering
\includegraphics[page=5, trim={0 0 15cm 3cm}, clip, width=0.9\linewidth]{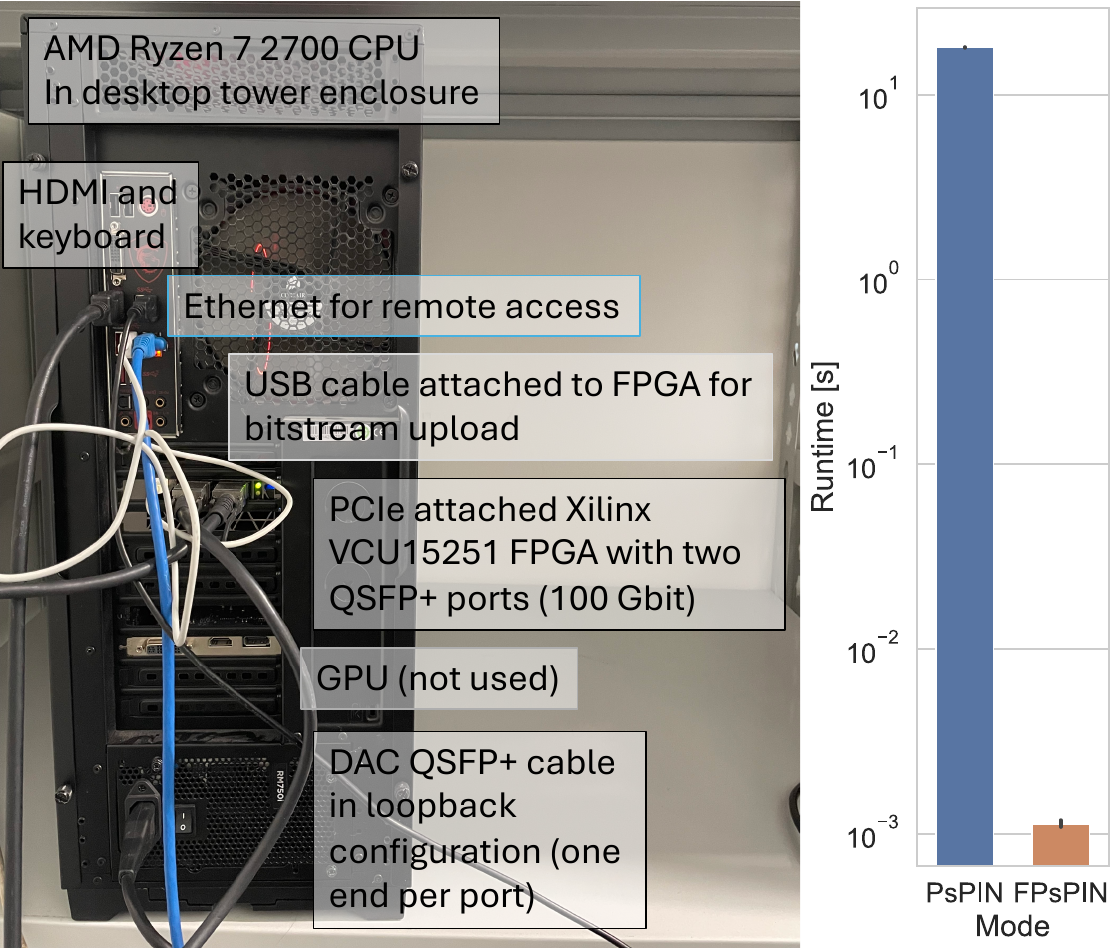}
\caption{Our FPsPIN prototype system and a performance comparison of this work and its predecessor PsPIN~\cite{pspin}.}
\label{fig:prototype}
\end{figure}

To this day, no full PsPIN-based sNIC has been demonstrated. While simulations are useful to show the \emph{potential} to process certain workloads at specific data-rates, only a full end-to-end implementation can ensure no usnimulated part of the system will lead to unforeseen slow-down.
Another drawback of cycle-accurate simulations is their long run time. In Figure~\ref{fig:prototype} we show a comparison between executing a 32 packet ping-pong on an FPGA and performing the same experiment as cycle-accurate simulation. Our work delivers a run time reduction of four orders of magnitude. 

We release our implementation~\cite{xu_thesis} 
to the public\footnote{\url{https://github.com/spcl/FPsPIN}} as an extensible platform for smart NIC research.
The main contribution of this work is an open-source FPGA-based full-system implementation of a PsPIN based sNIC, implementable using readily available hardware, an example system is shown in Figure~\ref{fig:prototype}. We present several use-cases of this implementation.

\section{Related Work}
\label{sec:relwork}

Modern NICs can be classified into three categories~\cite{wei2023characterizing}: RDMA capable NICs, on-path smart NICs and off-path smart NICs. RDMA capable NICs need to be able to steer data to the destination address, i.e., write to the correct address indicated within the packet and thus often have processing cores (for protocol processing) and DMA engines within them, but these cores do not need to be programmable by the user. On-path sNICs are architecturally similar, i.e., incoming data is processed by cores on the NIC and potentially DMAed to the host memory, usually via PCIe, the difference between an on-path sNIC and an RDMA-capable NIC (rNIC) is that the behaviour of the processing cores can be altered by the user. An off-path sNIC on the other hand contains two sets of cores, which are both located on the sNIC, the first enables the simple RDMA data-steering, the second is a System-on-Chip (SoC) connected internally, usually also via PCIe. Often off-path smart NICs run a full Linux operating system on the SoC and are thus easy to program by the user. The obvious drawback of off-path sNICs is the added latency incurred by crossing the internal PCIe links.

By the above classification, a NIC becomes a smart NIC as soon as the user is allowed to execute code on the NIC cores. Programmable NICs have existed in special-purpose environments. For
example, Quadrics QSNet~\cite{petrini2002quadrics} offered a programming interface that
was used to offload collectives~\cite{yu2004efficient}. However, QSNet had to be programmed at a very low level and was rather limited, which hindered wider adoption. Myrinet provided open firmware that allowed researchers to implement their own modules on the specialized NIC cores in C~\cite{wagner2004nic}.

As opposed to these vendor and platform specific solutions, sPIN
defines a CUDA-like offload interface for network operations --- handlers which are executed on incoming packets which match a certain pattern. Thus rewriting an application to make use of sPIN is hardware agnostic. This is similar to approaches that have been widely adopted for compute offloading.
High-speed packet processing frameworks used for router implementations, such as P4~\cite{bosshart2014p4} provide
similar abstractions, however, these frameworks are not designed to interact with host memory and the execution units are stateless and are thus less versatile than sPIN.

Instead of allowing users to execute code on NIC cores, several research projects have demonstrated the usefulness of an even lower-level abstraction: Using FPGA based NICs to allow users to add hardware blocks to the NIC~\cite{catapult, istvan2016consensus, forencich2020corundum}. While FPsPIN itself uses Corundum~\cite{forencich2020corundum} in its implementation, we fear that programming in a hardware-description language would be daunting for most users and lead to very use-case specific solutions and thus is no practical. Nevertheless, the open nature of the RISC-V instruction set used by the FPsPIN cores and the fact that we use an FPGA to implement them allows (but does not force) users to add instructions or other functionality accelerated by custom logic blocks.

Multiple vendors have built off-path sNICs in recent years, such as Broadcom Stingray~\cite{stingray}, NVIDIA Bluefield~\cite{bluefield}, and others~\cite{xing2022towards}. Their sNICs run full operating systems in order to offload data-processing tasks onto an SoC integrated into the NIC. While this frees CPU cycles, it does not change the number of times data has to be read from the NIC over PCIe into cache when processing and thus does not improve latency.

On the opposite end of the spectrum are frameworks such as DPDK~\cite{dpdk} which do not rely on sNICs at all but allow users to dedicate compute cores to packet processing and offer efficient drivers for NICs produced by multiple vendors to read and write network packets quickly from user-space.

\section{Background}

In this section we explain all the pieces that contribute to FPsPIN which have been developed outside of the scope of this work: the sPIN abstract machine model, the open-source sPIN implementation in Verilog PsPIN (which is lacking crucial NIC functionality such as a way to send and receive network packets) and the open-source NIC implementation Corundum which allows users to add custom hardware modules into the NIC.

\subsection{The sPIN Abstract Machine Model}
\label{sec:spin_model}

The acronym sPIN stand for \textbf{s}treaming \textbf{p}rocessing \textbf{i}n the \textbf{n}etwork~\cite{spin_orig}.
The underlying machine model is protocol agnostic, supports packet- as well as message-based protocols, and processes data on a packet granularity. A user of sPIN defines \emph{handlers}, small functions which are executed upon reception of packets into the NICs fast memory. The handlers are executed on handler processing units (HPUs) --- cores which have access to this fast memory. Packets are scheduled to HPUs by a packet scheduler, which contains a matching engine, which determines if a handler should be executed for a specific packet (if no such handler is defined, the packet is forwarded to the Corundum RDMA NIC). 

\begin{figure}[!htb]
\centering
\includegraphics[page=1,trim={2.5cm 2.5cm 5.5cm 3cm},clip, width=\linewidth]{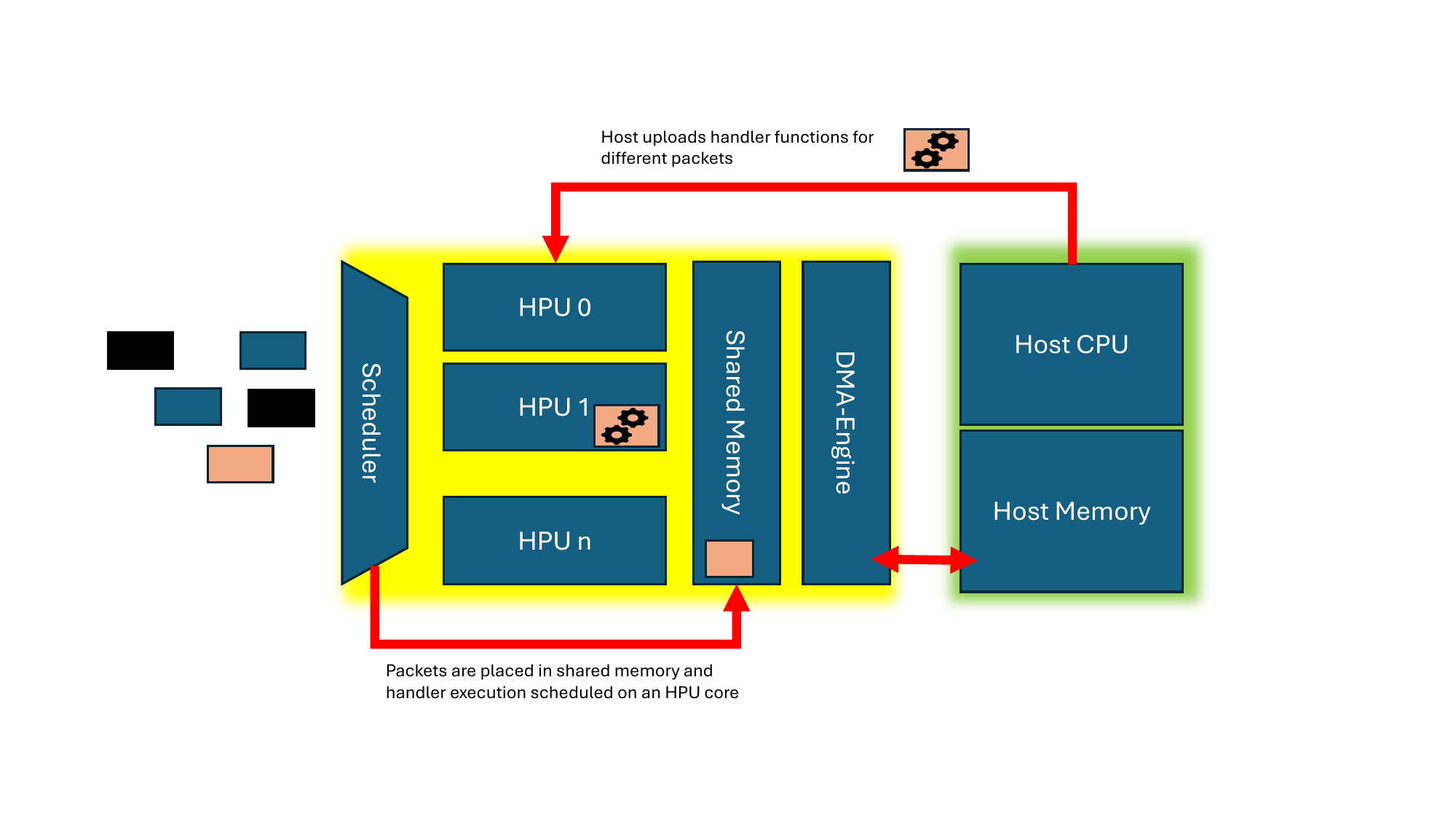}
\caption{The sPIN abstract machine model}
\label{fig:spin_model}
\end{figure}

HPUs also have access to a DMA engine which allows them to transfer data to and from host memory.
HPUs can also send packets via an outbound engine. Figure~\ref{fig:spin_model} gives an overview of the sPIN abstract machine model.

The handler types supported by sPIN are \emph{header-}, \emph{packet-} and \emph{tail-}handlers. The header handler is executed only on the first packet of a message, the tail handler only on the last. The definition of a message depends on the particular sPIN implementation, which is protocol and ISO/OSI layer~\cite{iso_osi_model} dependent, i.e., a sPIN implementation targeting a layer 2 protocol such as Ethernet might not define any kind of message, and simply execute the packet handler on every matching packet. A sPIN implementation targeting MPI on the other hand might define the first packet corresponding to an \texttt{MPI\_Send} as the beginning of a message and thus execute the header handler in addition to the packet handler when receiving such a packet.  On a network which guarantees in-order packet delivery, this allows the user to easily set up a context for processing a message in the header handler and destroy this context in the tail handler (which is executed for the last packet of a message).

Handlers are uploaded and removed to/from the sNIC using an \emph{execution context}: a rule which defines how a packet has to look like to be considered matching to a certain group of handlers, and the necessary context for the execution of the handlers, such as pointers to host-memory DMA regions. This allows the user to write a different handler for packets sent from a different source or corresponding to a different application. The handlers themselves are functions written in C. Each sPIN implementation defines how they are compiled.

The sPIN abstract machine model allows sPIN implementations to add additional hardware to their implementation, e.g., a deeper memory hierarchy with shared memory between HPUs or hardware modules for specific tasks such as computing protocol dependent packet checksums. While the usage of such implementation-dependent features limits the portability of sPIN-enabled programs, it also allows for vendor-specific extensions.

\subsection{FPGA}

An Field-Programmable Gate Array (FPGA)~\cite{fpga} is an integrated circuit that allows runtime reconfiguration. It consists of an array of configurable logic
blocks (CLB) that can function either as look-up table (LUT) or flip-flop (FF),
on-chip SRAM as block RAM (BRAM) or Ultra RAM (URAM) macros, and
programmable routing resources that connect the input and output of CLBs.
LUTs are used to implement combinatorial logic and FFs for sequential logic.
FPGA also contain high-speed transceivers for
high-speed buses, e.g. PCIe or DDR4. An FPGA is usually programmed using a hardware-description language (HDL), such as Verilog. The HDL source code is compiled into a bitstream, which, when flashed onto the device, configures all CLBs and routing resources to form a specific digital logic design. 

In order to modularize the design of digital logic designs, hardware
intellectual property (IP) vendors package function blocks
as so called IP blocks and redistribute them to customers for integration into their
design. For interoperability, IP blocks adopt standard bus protocols. On the other
hand, if design components do not speak the same bus protocol, an adapter is
needed. Adapters consume hardware resources and may impact performance; a perfect adapter may not even be possible in case of a mismatch of semantics between the protocols.

The Advanced eXtensible Interface (AXI) \cite{axi1, axi2} is a family of on-chip communication bus protocols designed to connect IP blocks in a hardware design. AXI
protocols follow a master-slave design where the bus master initiates transactions and the bus slave responds. The protocol has three flavours, designed
for different use cases. AXI-MM is designed for high-performance memory-mapped read and write access from processor-like masters on addressable memory-like slaves. AXI-Lite is designed for lightweight memory-mapped access for lower-performance situations; it does not support advanced features like bursting, narrow transfers or interleaved requests, making it a lot simpler to implement. AXI-Stream is designed for streaming data from master to slave without addressing semantics.

\subsection{The PsPIN Implementation of sPIN}

\begin{figure}[!htb]
\centering
\includegraphics[page=2, trim={0 7.5cm 7.6cm 0}, clip, width=\linewidth]{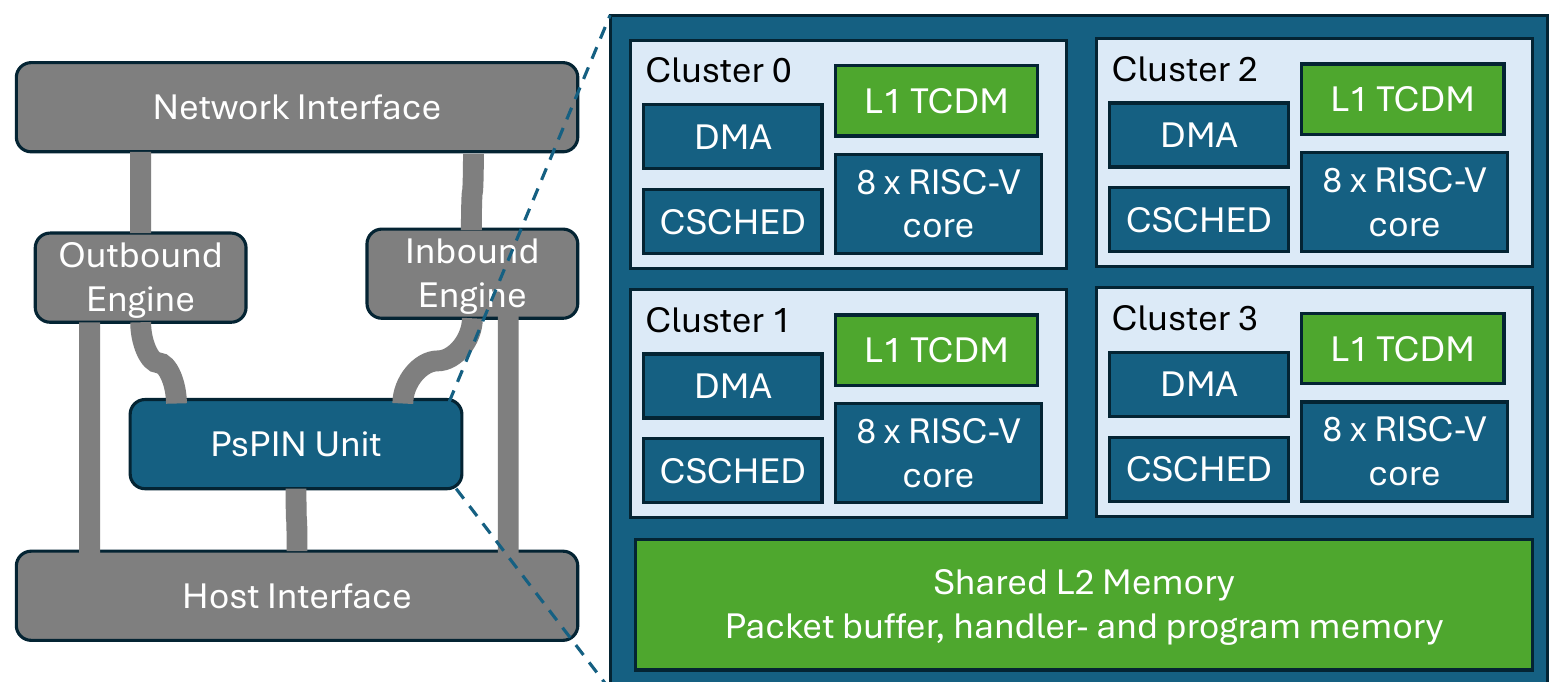}
\caption{The PULP sPIN architecture}
\label{fig:pspin}
\end{figure}

While sPIN defines an abstract machine model, it does not specify the exact micro-architecture of sPIN-enabled NICs such as the instruction set architecture (ISA) for the HPUs or the exact memory hierarchy on the NIC. PsPIN~\cite{pspin} is a open-source reference implementation of sPIN in Verilog, ready to be integrated into the packet data path of existing NIC designs. PsPIN uses the RISC-V~\cite{waterman2014risc} based CPU cores developed by the PULP~\cite{pulp} project to implement the HPUs.  It groups the HPU cores into clusters for a hierarchical memory architecture and multi-level scheduling. An overview of the PsPIN architecture and NIC model is shown in Figure~\ref{fig:pspin}.

Note that the PULP architecture is configurable, i.e., the number of clusters and cores per cluster can be changed. We make use of this in our FPGA implementation in order to maximise the usage of the FPGA platform while at the same time controlling for the length of the critical path. 

The control flow of PsPIN starts with new packets arriving from the NIC. The NIC inbound engine generates a handler execution request (HER) that contains metadata for scheduling the packet on a HPU, including the address of the packet data in the NIC memory and the address of the sPIN handler functions. The packet scheduler then resolves the scheduling dependencies, e.g., it ensures header-handlers are scheduled before packet-handlers, according to the sPIN abstract machine model as described in Section~\ref{sec:spin_model} and forwards individual tasks to the cluster schedulers. The cluster scheduler forwards the incoming tasks to the HPUs for execution, and collects the completion notifications from the HPUs. It forwards the completion notifications through the global scheduler back to the NIC inbound engine, such that the packet buffer can be de-allocated and reused.

Three major data flows cover the full cycle of packet processing in PsPIN, all of which are driven by various DMA engines, allowing fast data movement and latency hiding. The inbound packet data from the NIC inbound engine to the L2 packet buffer is handled by the DMA engine in the NIC inbound engine; the data is further DMA’ed into the cluster-local L1 memory by the cluster DMA engine. Host memory access by the HPUs flow from L2 or L1 to the host memory and is handled by the off-cluster DMA engine. Outgoing packet data from L2 or L1 is handled by the DMA engine inside the NIC outbound engine. PsPIN exposes AXI slave ports for access to the internal interconnect by the NIC DMA engines.

\subsection{Corundum}

\begin{figure}[!htb]
\centering
\includegraphics[width=\linewidth,height=1in]{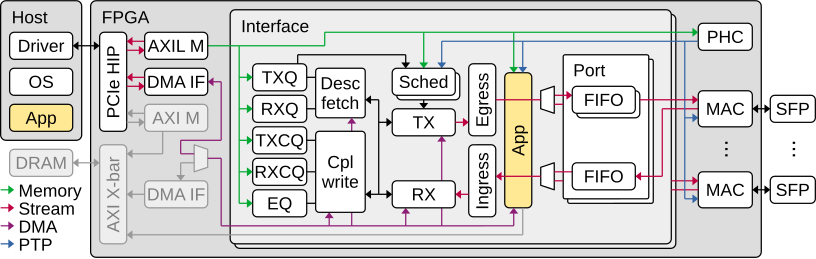}
\caption{Corundum architecture overview from the corundum documentation~\cite{corundum_doc}, the app block, in which FPsPIN is implemented, is highlighted in yellow.}
\label{fig:corundum}
\end{figure}

Corundum~\cite{forencich2020corundum} is an open-source, FPGA-based rNIC. It supports 10/25/100 Gbps Ethernet on Xilinx and Intel FPGA platforms. It offers a high-performance, PCIe DMA system and open-source platform-agnostic IPs including the Ethernet media access control (MAC) layer and AXI infrastructure. Corundum offers a full software stack on Linux. 
Corundum is amenable to custom extensions to broaden the functionality of the rNIC. Such extensions are foreseen as the \emph{App} block, highlighted in yellow in Figure~\ref{fig:corundum} (taken from the Corundum documentation).

The App block has access to the control path, data path, and DMA subsystems. Corundum’s software stack also provides kernel interfaces for custom drivers and user space utilities.

\section{The FPsPIN FPGA Prototype}
\label{sec:fpspin_arch}

PsPIN itself is not a fully functional SmartNIC due to the lack of
capability to send and receive packets and the in-ability to access host memory.
FPsPIN combines PsPIN and Corundum to form a complete sNIC.

\begin{figure}[!htb]
\centering
\includegraphics[page=3, width=\linewidth]{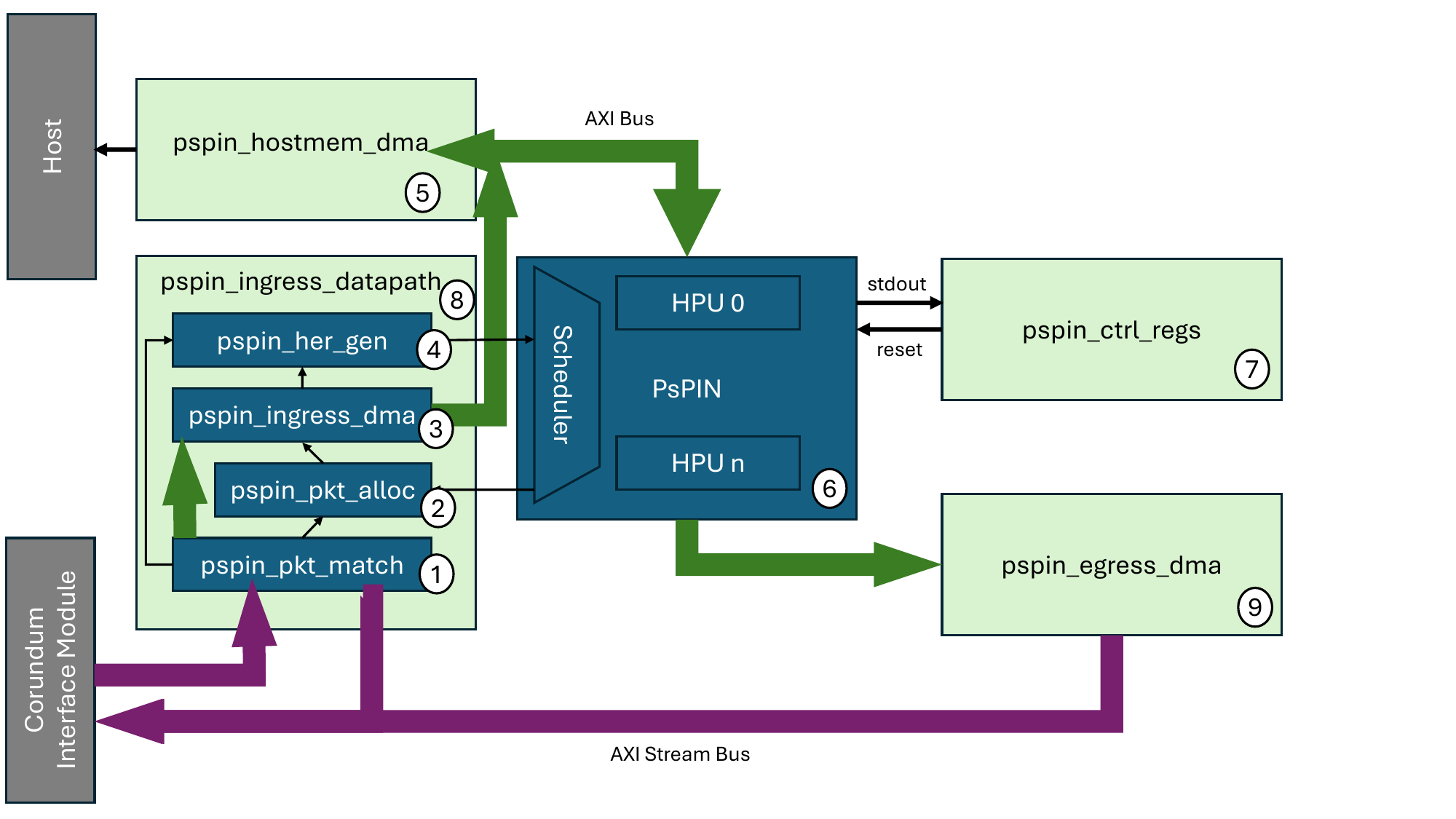}
\caption{FPsPIN architecture overview}
\label{fig:fpspin_block}
\end{figure}

An overview of all the FPsPIN hardware components is shown in Figure~\ref{fig:fpspin_block}, the four green
blocks have been added to PsPIN \Circled[]{6} by this work:
The ingress \Circled[]{8} and egress \Circled[]{9} data paths allow the PsPIN cluster to receive packet data from the network and send packets into it. The control path \Circled[]{7} allows the PsPIN cluster and other components to be configured from the host over various control registers and program memory. The host-side DMA \Circled[]{5} enables the PsPIN cluster to read from and write to the main memory on the host system.

Upon arrival of a packet from the Corundum ingress engine, we first check if this 
packet should be handled by the sNIC cores, i.e., if there is currently a handler loaded for this type of packet; this is done by the \textbf{pspin\_pkt\_match} block \Circled[]{1}. If this is not the case, the packet is forwarded to the Corundum data path. The matching rules in our implementation are similar to those implemented by the iptables~\cite{hoffman2003testing} U32 matcher~\cite{iptables_u32}: The user supplies an index I, start and end values S and E respectively, as well as a mask M. A rule matches if the 32-bit value of packet data at byte-index 4I:4I+3 masked with M lies between S and M. Applying such a matching rule to a packet emits a single boolean output, and we allow 
to combine three such rules either via an AND or OR operation.  An example for setting up a matching rule for ICMP Echo-requests is shown in Section~\ref{sec:writing_handlers}. These matching rules allow FPsPIN to ignore traffic which belongs to protocols not handled by FPsPIN handlers, 
for example the the Address Resolution Protocol (ARP) is essential for the correct functioning of an IP network (it establishes the mapping between layer 2 MAC addresses and layer 3 IP addresses). Since non-matching packets are forwarded to the Corundum datapath, ARP traffic will still reach the Linux kernel network stack (assuming the user did not plan on accelerating the ARP protocol itself by implementing handlers for it). This property makes FPsPIN easy to integrate into an existing network, since it behaves like a ``standard'' NIC, unless instructed otherwise for a specific traffic class.

The packet buffer allocator \Circled[]{2} \textbf{pspin\_pkt\_alloc}  allocates a buffer for the packet in the L2 memory packet
buffer of PsPIN if the packet is supposed to be processed by PsPIN. It 
adds the address of the allocated buffer to the packet metadata,
and forwards the metadata to the DMA engine to actually write the packet
into the memory. It takes in the completion notifications from PsPIN, to
determine when a packet has been processed and its buffer can be freed. 
The verilator model originally developed in the PsPIN project uses a software-based ring buffer in the simulation testbench to allocate space for incoming
packets in the packet buffer. This forces the allocator to keep a list of
out-of-order frees and is thus difficult to implement in hardware. However,
most packets on the Internet and in data center environments follow a
bimodal distribution in size with 40\% of packets are below 64 bytes and another
40\% are 1500 bytes~\cite{john2007analysis, benson2010understanding}. We thus take a simpler fixed-size allocation approach. We partition the packet buffer into two halves; in one half we make fixed 128-byte slots, and in the other half we make 1536-byte slots, and store free slots
in two separate FIFOs. We then handle allocation and free simply by popping
from and pushing to the respective FIFOs.

The ingress DMA module \textbf{pspin\_ingress\_dma} \Circled[]{3} takes the allocated address and
length in the packet metadata and performs a DMA transaction to write
the packet data to the PsPIN NIC inbound memory port. Upon finish of
the DMA request, the module forwards the packet metadata on to the HER
generator in the data path. We use the \emph{axi\_dma\_wr} module from the Corundum AXI IP
library to perform the actual DMA operation.

Once the transfer of the packet into the PsPIN L2 memory packet buffer is complete, the
\textbf{pspin\_her\_gen} module \Circled[]{4} issues a HER to PsPIN. Part of the information required to generate a HER comes from the packet metadata, such as the message ID and if the packet is
the last in a message (tail-handler needs to be executed). The rest of the HER stores the
address of the handler functions that the packet should be processed with, as
well as the host DMA regions.

The \textbf{pspin\_egress\_dma} module \Circled[]{9} handles egress commands from PsPIN: using the
Corundum IP \emph{axi\_dma\_rd}, the module performs a DMA read from the packet
buffer and gets an AXI-Stream bus that contains packet data. It then injects
the AXI-Stream into the outbound AXI Stream of Corundum using the AXI-Stream arbiter \emph{axis\_arb\_mux}. 

Access to host memory from FPsPIN is implemented within the module \textbf{pspin\_hostmem\_dma} \Circled[]{5}. It bridges the AXI master port of the PsPIN cluster to
the segmented DMA interface of Corundum, which takes a RAM port
and a separate command bus. We utilize the AXI-Stream DMA client (\emph{dma\_client\_axis\_source}, \emph{dma\_client\_axis\_sink}) from Corundum to convert the
output AXI Stream bus to AXI4 channels. For write requests from PsPIN,
the module first issues a DMA command to the AXI-Stream client to capture
the write data in a dual-port RAM buffer; it then issues a
command to the Corundum DMA subsystem to DMA the data from the
buffer RAM to the host memory. The read process happens in the reverse
order. The adapter is not fully AXI-compliant: we do not support irregular bursts (narrow bursts or modes other than INCR), as well as interleaved
read requests. Unlike AXI4, the PCIe interface also does not support arbitrary
byte enable (BE) configurations, so we also do not support these cases.
However, these limitations are acceptable in our use case, since the DMA bus master in PsPIN does not issue such requests.
An important feature we do implement are unaligned
writes, which are essential to some applications such as MPI derived datatype processing.
While it is possible to implement unaligned transfers in software by reading
the affected word first to compose and issue an aligned transfer, the extra
memory read transactions (up to two extra reads for one unaligned write)
would significantly hurt performance. Fortunately, the Corundum DMA
subsystem fully supports unaligned transfers. As AXI4 expresses unaligned
writes as aligned writes with strobe (byte-enable), we implement an address
recovery procedure that calculates the original address and length from the
AXI burst strobe (WSTRB) signal of the first and last beat in the AXI transaction.
The module then issues the unaligned transfer to the client and Corundum
DMA subsystem.

\subsection{Specialization of PsPIN for FPGA Implementation}

PsPIN uses the PULP Ultra Low Power (PULP) RISC-V cores
and AXI infrastructure, which are originally designed for ultra-low-power
ASIC platforms. This means that they are optimised for recent ASIC process
nodes and thus have long critical paths, which are suboptimal for FPGA
implementation. While some parameter tweaking allowed us to break very long
critical paths e.g. single cycle bus across the entire SoC, most components
need to be redesigned to reach a higher clock speed on FPGAs. 

\begin{table}[!htb]
    \centering
    \begin{tabularx}{\columnwidth}{lXX} \hline
    Resource Category & PsPIN~\cite{pspin} & FPsPIN \\ \hline
    \# of Clusters & 4 & 2 \\
    \# of MPQ Entries & 256 & 16 \\
    L1 Cluster Memory & 1 MiB & 256 KiB \\
    L2 Program Memory & 32 KiB & 32 KiB \\
    L2 Packet Memory & 4 MiB & 512 KiB \\
    L2 Handler Memory & 4 MiB & 1 MiB \\ \hline
    \end{tabularx}
    \caption{Comparison between the stock PsPIN configuration and that used in FPsPIN.} \label{tab:pspin-config}
\end{table}

\begin{table}[htb]
    \centering
    \begin{tabularx}{\columnwidth}{llXX} \hline
    Module & Cycles & Frequency (MHz) & Latency (ns) \\ \hline
    Matching engine & 4 & 40 & 100 \\
    Allocator & 0 & 40 & 0 \\
    Ingress DMA & 8-70 & 40 & 200-1750 \\
    HER generator & 0 & 40 & 0 \\
    Host DMA & \emph{n/a} & 250 & 450 \\ \hline
    \end{tabularx}
    \caption{{Latency estimation for various data path modules in cycles and nanoseconds.}} \label{tab:lat-cycles}
\end{table}

The lengthy critical paths of PULP and thus PsPIN on FPGAs mean that
without significant re-engineering, the packet processing cluster could only
run at a lower frequency. This situation is further worsened by the area
requirements of the original PsPIN design: the 4-cluster configuration that
was used in PsPIN~\cite{pspin} proved to be extremely difficult, to place and route on the used FPGA device. We thus
use a 2-cluster configuration with reduced memory sizes. To further resolve
the routing congestion problems, we employ the incremental implementation
flow.
In contrary to PsPIN, Corundum runs at 250 MHz on the target Xilinx VCU15251 Development Kit
device, shown in Figure~\ref{fig:prototype}. While it is possible to retarget Corundum to run at a lower frequency,
we would have to reconfigure the clock domains and validate that the
resulting design still works properly; Instead,
we opted to only run the PsPIN cluster and the closely coupled data path
engines at 40 MHz for the evaluation.

\begin{table}[!htb]
    \centering
    \begin{tabularx}{\columnwidth}{XXX} \hline
    FPGA Resource & Used & \% of avail \\  \hline
    LUT & 645k & 54.5\% \\
    FF & 490k & 20.7\% \\
    BRAM & 1141 & 52.8\% \\
    URAM & 206 & 21.5\% \\
    Implementation Time & 6h:11m & n/a \\ \hline
    \end{tabularx}
    \caption{Resource usage of the hardware implementation of FPsPIN on the VU9P device.}
    \label{tab:design-qor}
\end{table}

To evaluate the implementation of the
components introduced in Section~\ref{sec:fpspin_arch} we estimate the theoretical latency of these components based on the RTL source. Table~\ref{tab:lat-cycles} shows the latency
in cycles based on the state machine construction in the Verilog code, the frequency the respective module is running at, and the resulting latency in nanoseconds.

Resource utilisation and timing are very important static insights into FPGA designs.  We had to trim the original PsPIN design significantly compared to the standard configuration~\cite{pspin} as shown in Table \ref{tab:pspin-config}, in order to close timing.   To ensure that we get acceptable implementation results for each run, we employ the \emph{incremental implementation flow} from Xilinx to have the EDA tool try to reuse routed nets from previous valid implementation runs.

We use Xilinx Vivado 2020.2 EDA suite to produce the FPGA bitstream. We show the resource utilization on our target platform in Table~\ref{tab:design-qor}. We note that our FPsPIN design uses only half of most available resources, this is a consequence of the many busses involved, which have to be routed accross the FPGA. An interesting research direction would be to replace the PULP cores used in this work as HPU cores with cores that are specifically designed for FPGAs. However, as one of the goals for this work was to validate the existing PsPIN design, we consider that future work.

\subsection{FPsPIN Software Stack}

The FPsPIN software stack consists of three equally important parts, one in kernel-space and a user-space API which allows users to make use of sPIN. The third is the sPIN runtime which is used by sPIN handlers. We describe all three in this section in the previously mentioned order.

FPsPIN comes with a dedicated kernel module that interfaces FPsPIN with
existing subsystems in Linux. Corundum ships a kernel driver for the complete NIC functionalities, including interactions with the Linux network stack to expose the device as
Ethernet NIC for packet transmit and receive, as well as control interfaces. The Corundum driver registers the application block as an auxiliary device and exposes the application base address (a separate PCIe BAR) to the auxiliary driver. This allows the custom driver to access the
application block BAR to interact with the hardware. The FPsPIN kernel module has to fulfil various high-level tasks: Allowing users to \emph{debug} sPIN handlers, i.e., forward debug output from handlers to userspace, allow \emph{access to PsPIN memory} from the host in order to upload handlers, and \emph{facilitate DMA} from the PsPIN clusters to the host memory, i.e., making sure pages which can be the target of a DMA write are not swapped out.

To allow \textbf{debugging} of handlers the HPUs write their standard output into a FIFO for the host CPU to read for diagnostic purposes. The FPsPIN kernel driver exposes the \texttt{/dev/pspin1}
character device to the user-space, which can be polled and delivers multiplexed output from all HPUs.

The kernel module \textbf{exposes the PsPIN memory} in two ways: The first is via
traditional non-buffered file I/O using the open(), seek(), read(), write(), and close() syscalls on the \texttt{/dev/pspin0} character device.
Reads and writes to the device file are translated into reads and
writes in the NIC memory region. 
The second way to access PsPIN memory is via 
an ioctl() (a system call whichs semantics depend on the device it is applied to and the request argument supplied with that system call) on the \texttt{/dev/pspin0} device
file. It allows the user-space application to pass a pointer, a 64-bit word and a direction (read or write) to allow faster (single system call as opposed to three) access to single values.

Memory pages used for \textbf{DMA} have to be registered
before usage e.g., to make sure that the memory page used for
DMA are not moved by the kernel through swapping or memory compaction.
To this end, the Linux kernel offers a DMA API. We implement the mmap()
syscall for the \texttt{/dev/pspin0} device to perform a multi-use DMA allocation.
We then mark the area as uncached and map the
allocated DMA memory area into the user application address space.
The user-space needs access to the physical address of the mapped
DMA area to write to the control registers. We implement another ioctl on
\texttt{/dev/pspin0}, to allow the user-space to query the
physical address of the DMA area, in order to set up the execution context.

\subsection{Writing FPsPIN Handlers}
\label{sec:writing_handlers}

When using FPsPIN, the user writes up to three functions (header-, tail-, and packet-handler) which are compiled for the PsPIN cores, using the PULP software SDK~\cite{pulp_sdk}. In Listing~\ref{lst:udp_ping_pong} we show the packet-handler for a UDP ping-pong server which for each packet received, simply returns the packet to the sender. Since the UDP protocol is stateless, there is no need to set up or tear down any context in a header- or tail-handler.

\begin{pyglist}[language=c, fontsize=\footnotesize,  numbers=left,numbersep=2pt, caption={A FPsPIN packet handler which responds to UDP Ping Packets}, label={lst:udp_ping_pong} ]
#include <pspin.h>
#include <handler.h>
#include <packets.h>
#include <spin_dma.h>

__handler__ void ping_ph(handler_args_t *args) {
  task_t *task = args->task;
  pkt_hdr_t *hdrs = (pkt_hdr_t *)(task->pkt_mem);
  uint16_t pkt_len = args->task->pkt_mem_size;
  
  // swap ETH src and dst MAC address
  mac_addr_t src_mac = hdrs->eth_hdr.src;
  hdrs->eth_hdr.src = hdrs->eth_hdr.dest;
  hdrs->eth_hdr.dest = src_mac;

  // swap src and dst address in IP header
  uint32_t src_id = hdrs->ip_hdr.source_id;
  hdrs->ip_hdr.source_id = hdrs->ip_hdr.dest_id;
  hdrs->ip_hdr.dest_id = src_id;

  // swap src and dst port in UDP header
  uint16_t src_port = hdrs->udp_hdr.src_port;
  hdrs->udp_hdr.src_port = hdrs->udp_hdr.dst_port;
  hdrs->udp_hdr.dst_port = src_port;

  spin_cmd_t put; //sPIN commands are non-blocking
  spin_send_packet(task->pkt_mem, pkt_len, &put);
  spin_cmd_wait(put); // wait for send to complete
}
\end{pyglist}

Each handler function invocation gets passed a \texttt{handler\_args\_t} pointer. The two most important members of this type are \texttt{task->pkt\_mem}, a pointer to the packet the handler is called for in the PsPIN L1 memory and \texttt{task->pkt\_mem\_size}, the length of the packet in bytes. For this simple task we can ignore other members. We use a convenient header-file supplied by FPsPIN which defines C structures for commonly used protocol headers in order to swap the source and destination ethernet MAC, the IP address and the UDP source and destination port. We then send the packet. Note that most FPsPIN functions which move data are non-blocking, thus we have to explicitly wait for the command to complete before we can return from the handler.

To install the handler and react to incoming packets we have to set up an execution context and upload our compiled code to a sPIN-capable NIC. Listing~\ref{lst:ping_pong_host} shows code to load a handler that implements an ICMP Echo server. 

\begin{pyglist}[language=c, fontsize=\footnotesize, caption={Loading an FPsPIN handler which responds to ECMP Echo Requests.}, label={lst:ping_pong_host}, numbers=left, style=tango, numbersep=2pt]
#include "fpspin/fpspin.h"
#include <unistd.h>

void ruleset_icmp_echo(fpspin_ruleset_t *rs) {
  *rs = (fpspin_ruleset_t){
    .mode = FPSPIN_MODE_AND,
    .r = {
       FPSPIN_RULE_IP,
       FPSPIN_RULE_IP_PROTO(1), // ICMP
       ((struct fpspin_rule){
         .idx = 8,
         .mask = 0xff00,
         .start = 0x0800,
         .end = 0x0800}), // ICMP Echo-Request
       FPSPIN_RULE_FALSE,  // no EOM
    }, };
}

int main(int argc, char *argv[]) {
  fpspin_ctx_t ctx;
  fpspin_ruleset_t rs;
  ruleset_icmp_echo(&rs);
  fpspin_init(&ctx, "/dev/pspin0", "build/pingpong", 
    0/*dst ctxt*/, &rs, 1/*dst ruleset*/,
    FPSPIN_HOSTDMA_PAGES_DEFAULT);
  while (1) {sleep(10);}
  fpspin_exit(&ctx);
}
\end{pyglist}
The reason why we chose this example is because it demonstrates the FPsPIN packet matching mechanism in more detail. In this example we want to match only ICMP echo request packets.

\begin{figure}[!h]
\centering
\includegraphics[page=4, trim={0 14cm 5cm 0}, clip, width=\linewidth]{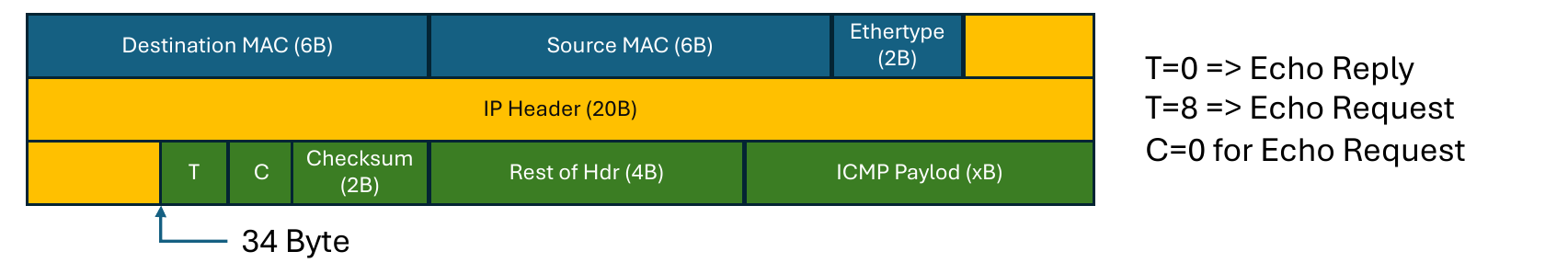}
\caption{Anatomy of an ICMP Packet - to match ECMP Echo Requests we need to check if Byte 34 has a value of 8.}
\label{fig:icmp_hdr}
\end{figure}

Figure~\ref{fig:icmp_hdr} shows the various headers of an ICMP packet as it arrives to the FPsPIN matching unit, the Ethernet header is marked in blue, the IP header is marked in yellow, and the ICMP header in green. An ICMP Echo-Request is an IP packet, where the Identification field  of the IP header contains the value 1, and the value of the Type field (abbreviated as T in the Figure) in the ICMP header is 8. FPsPIN comes with predefined rules to match an IP packet (\texttt{FPSPIN\_RULE\_IP}) and IP protocols by number (\texttt{FPSPIN\_RULE\_IP\_PROTO(n)}). For ICMP Echo Requests we define a custom rule: As shown in Figure~\ref{fig:icmp_hdr} we need to ensure byte 34 of the packet has a value of 8. Since the FPsPIN matching engine matches on 32-bit words, we supply an index of 8 to select bytes 32:35.
We then use a mask of \texttt{0xff00} to mask out everything except
the first two bytes (packets are transmitted in network byte order). The start and end values of \texttt{0x0800} ensure that all packets which match this rule have a value of 8 in that position. The mode of \texttt{FPSPIN\_MODE\_AND} ensures all three supplied rules (IP, IP protocol type 1, ECMP Echo-Request) have to match in order for a packet to be processed by our handler. The last supplied rule in each context has a different function, it identifies packets that mark the end of a message. In the ICMP server case above we are not making use of header- or tail handlers, so we simply use the predefined rule \texttt{FPSPIN\_RULE\_FALSE} which never matches.
One shortcoming of the current matching engine implemented in FPsPIN is its inability to deal with packets with a variable header length, however, if that is required a user can match all packets and perform filtering in the handler itself. 

While the above is a minimal working example, Table~\ref{tbl:spin_funcs} gives an overview of currently implemented functionality in FPsPIN: Apart from the ability to send packets and DMA memory to and from the host, the sPIN runtime also offers support for various types of locks, performance measurements and a simple to use FIFO queue for host communication.

\begin{table}[!htp]
    \centering
    \begin{tabularx}{\columnwidth}{lX} \hline
    Function & Semantics\\  \hline
     spin\_send\_packet(ptr, len, cmd) & Send a single Ethernet packet \\
     spin\_dma(src, dst, len, dir, opts, cmd) & DMA from NIC to host or vice-versa \\
     spin\_write\_to\_host(..) & Similar to DMA for 64-bit words, blocking \\
     spin\_dma\_wait(cmd) & Blockingly wait for DMA completion \\
     spin\_dma\_test(cmd, flag) & Check for DMA completion \\
     spin\_cmd\_wait(cmd)      & Wait for the completion of a non-blocking sPIN function \\
     spin\_lock\_init(lck) & Init a lock \\
     spin\_lock\_lock(lck) & Block until lock is granted \\
     spin\_lock\_unlock(lck) & Unlock lck \\
     spin\_lock\_try\_lock(lck) & Check if lock is available \\
     spin\_rw\_lock\_*(lck) & Reader-writer lock, similar to standard lock but allows multiple readers to hold lck \\
     cycles() & Get the cycle-granular timestamp \\
     push\_counter(queue, val) & enqueue a value into a FIFO readable from host \\ 
     \hline
    \end{tabularx}
    \caption{Functions provided by the FPsPIN handler runtime and library}
    \label{tbl:spin_funcs}
\end{table}

\section{Demonstration of FPsPIN}

The experiments are done on the AMD server with the Ryzen 7 2700 CPU and
a PCIe-attached Xilinx VCU15251 Development Kit. The FPGA board uses 16 lanes of PCIe 3.0 clocked
at 8 GT/s. Corundum runs at its native frequency of 250 MHz, the application block (FPsPIN) at 40
MHz. The two 100 Gbps QSFP Ethernet ports on the FPGA board
are attached via one direct-attached copper (DAC) cable, forming a loop-back
between the two interfaces of Corundum. Since the two interfaces are present
on the same Linux host, we isolate the two network interfaces
using network namespaces to avoid a direct loop-back in software.
We use Ubuntu 20.04.4 on the host with a
Linux 5.15.0-76-generic kernel with the Contiguous Memory Allocator (CMA)~\cite{nazarewicz2019deep}
enabled; this allows the FPsPIN kernel driver to allocate arbitrarily large
contiguous DMA areas.

\subsection{Ping-Pong}

We demonstrate the overall system functionality with two classic types of ping-pong protocols: Internet Control Message Protocol (ICMP) and UDP. An important difference between both implementations is that ICMP requires the entire payload to be included in the  checksum calculation, while UDP only specifies an optional checksum of the UDP header which we omit. Using our FPsPIN prototype we implement three different versions of each ping-pong test: In \textbf{Host} mode the FPsPIN matcher is configured to never match, thus all packets are simply forwarded to the host and standard system utilities (ping for ICMP and dgping for UDP) are used to implement the ping-pong benchmark. In \textbf{FPsPIN} mode the matching unit matches all ping-pong related packets, performs all checksum calculations and sends the response to the client. The host CPU is not involved in this case. We can also exercise the host-DMA datapath of FPsPIN in \textbf{Host+FPsPIN} mode, in which FPsPIN matches all packets, forwards the received data to the host, which performs the checksum calculation, DMAs the result back to PsPIN, which then sends the reply.

\begin{figure}[!htb]
\centering
\includegraphics[width=\linewidth]{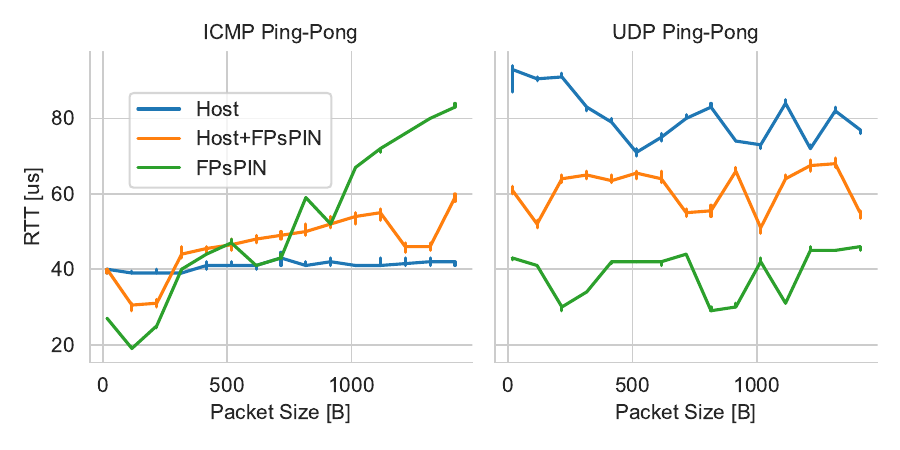}
\caption{ICMP and UDP ping-pong benchmark, using only the host, only FPsPIN or both.}
\label{fig:ping_pong}
\end{figure}

In Figure~\ref{fig:ping_pong} we show the median RTT for 20 measurements per scenario and data-size, together with the 95\% confidence interval (small bars). 
Both FPsPIN and Host+FPsPIN performed significantly better than Host for UDP, which demonstrates the large overhead of traversing the Linux UDP network stack, and context
switch to user-mode to reach the UDP responder. The ICMP responder in
Host, in comparison, runs in the Linux kernel and thus does not have the
overhead from the full UDP stack and context-switch to user-mode. This
overhead can be confirmed with a comparison between UDP and ICMP in
Host mode, showing a difference of 40 us.

We notice a big divergence in RTT with respect to payload size between ICMP and UDP: in the two modes that involve FPsPIN for ICMP, the RTT increases  linearly with the payload size, while in
the Host mode for ICMP as well as all three modes for UDP, the RTT remains
relatively constant. This reflects the difference in checksum calculation between the ICMP and UDP ping protocols, showing that checksum calculation
time is a significant component of the ICMP response RTT. The lower frequency of the FPsPIN HPU cores have a significant impact on packet processing latency. However, a single core on FPsPIN is only 2x slower than a single CPU core, a gap way smaller than the actual clock speed difference between these cores (40 MHz vs 3.4 GHz). In addition we must note that we used the identical portable C code to compute the ICMP checksum both in the FPsPIN ICMP handler and the host part of the ICMP Host+FPsPIN experiment, which shows a linear slowdown as well, the checksum calculation code in the Linux kernel on the other hand is highly optimized. The increase of RTT with respect to packet size is not caused by the DMA transfer to the host, otherwise we would see the same effect in the UDP Host+FPsPIN scenario.

\subsection{Reliable File Transfer}

The sPIN network offload model does not impose an underlying network
protocol; instead, it specifies two matching modes: In packet matching, single packets are matched for processing on the packet handler in the same flow; Message matching, on the other hand, requires the network to provide an abstraction of messages as a stream of multiple packets;
Examples of a network operating in message matching mode are RDMA-style
networks such as InfiniBand~\cite{infiniband} or the Intel Omni-Path Architecture~\cite{infiniband}.

Although FPsPIN is built on Ethernet, which would seemingly force a packet
matching implementation, many applications still benefit from the message-oriented abstraction sPIN offers, it is thus desirable to
emulate the message abstraction on top of Ethernet. Ethernet
is a lossy network (RDMA over Converged Ethernet (RoCE)~\cite{roce} does provide a lossless guarantee on top of Ethernet, but it is not supported by Corundum at the time) but traditional HPC middleware such as MPI require reliable message delivery. Thus we developed a protocol which can be fully implemented within sPIN handlers and provides reliability and flow-control. The protocol uses a 10-byte header inside the UDP payload with the following three fields: 
\textbf{Flags} (2 bytes): three packet-level status bits, synchronisation
(SYN), acknowledgement (ACK), and end-of-message (EOM); \textbf{Message ID} (4 bytes): unique ID of the message; and
\textbf{Offset} (4 bytes): byte offset of the first payload byte in the message.

\begin{figure}[!htb]
\centering
\includegraphics[width=\linewidth]{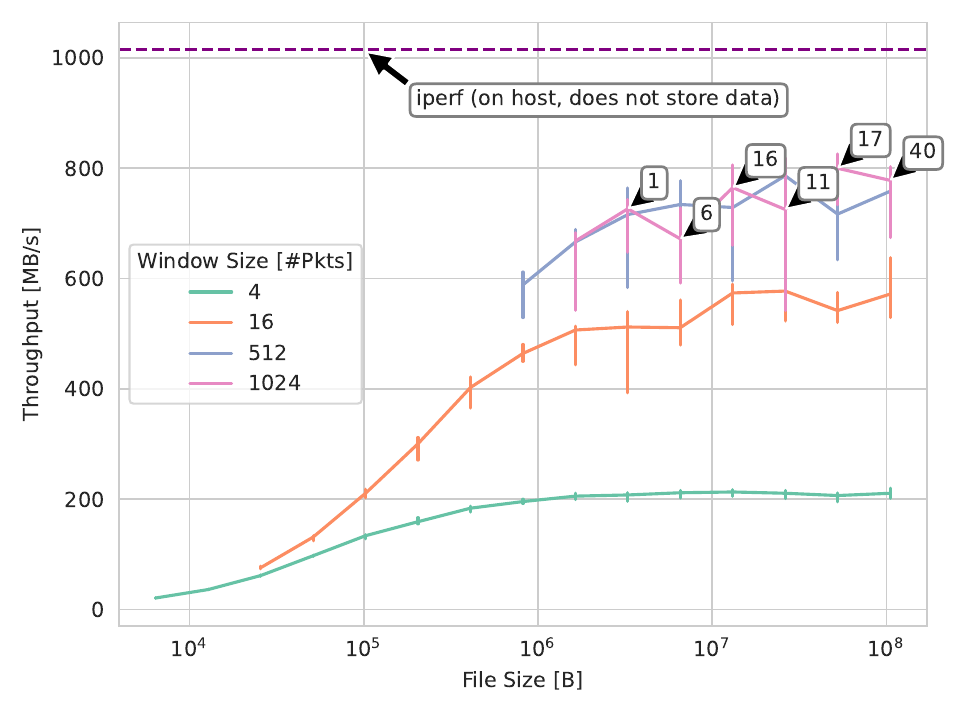}
\caption{Throughput of SLMP file-transfer using different window-sizes, compared to IPerf3 as a baseline. The annotations show how many SLMP segments could not be transmitted correctly (we ran each experiment until 20 successful transmits occurred and plot their median throughput and 95\% confidence interval).}
\label{fig:throughput}
\end{figure}

The SYN and ACK bits in the flags field in the SLMP header enable the sender to
decide in which reliability mode the SLMP protocol operates in. The receiver
receiver has to ACK each packet with the SYN bit set by sending back the same header with no payload.
For no reliability guarantee at all, the sender omits the SYN bit for all packets. For a guarantee of message
delivery but not individual segments, the sender sets SYN on the first and last
packets of the message. Since ACK's carry the Message ID and Offset, the sender is able to identify lost segments.

If the sender transmit packets too fast to the receiver, the receiver
will be overwhelmed by incoming packets and does not have time to process them; packets would be dropped once the receive buffer is completely filled.
Flow control throttles the sender to make sure that the receiver is not overwhelmed.
The SLMP protocol allows the sender to implement different forms of flow-control: either by
adjusting the inter-packet gap and thus the overall rate at which packets are sent until no packets are lost, or by choosing a threshold number of packets which are sent until the sender wait for an ACK (comparable to TCP windows). An interesting side-effect is that by forcing an ACK on every single packet of a message, i.e., using a window of size one, a sender can ensure packets are processed in-order by the receiver (we utilize this property when offloading MPI Datatype processing).

In Figure~\ref{fig:throughput} we show the throughput achieved when transferring files of different sizes using the offloaded SLMP protocol. For each data-point we show the median throughput of 20 successful (no segment lost) transfers and the 95\% confidence interval. File contents are DMA'ed to the host. Unsurprisingly, a larger window-size leads to a higher throughput, however, increasing the window-size even further does not provide any benefit, as almost all transfers fail, as shown by the text annotations which show the number of failed transfers we observed until 20 transfers completed successfully. As a comparison, we plot the median steady-state throughput of 20 runs of iperf~\cite{iperf} (runs on the host and does not use FPsPIN), a standard network benchmark which transmits UDP packets and then drops them upon reception. Our FPsPIN handlers (which write the received data to memory) achieve up to 80\% of the iperf baseline. In this example we use two sending threads (we were unable to saturate the receiver bandwidth with a single thread) which only send MTU sized packets and full SLMP windows. These restrictions are the reason for the ``missing'' data on the left-hand side of the plot: if MTU $\times$ window size $\times$ 2 exceeds the file size, the experiment was not performed.

\subsection{MPI Datatype Processing}

\begin{figure}[!htb]
\centering
\includegraphics[width=\linewidth,height=2.5in]{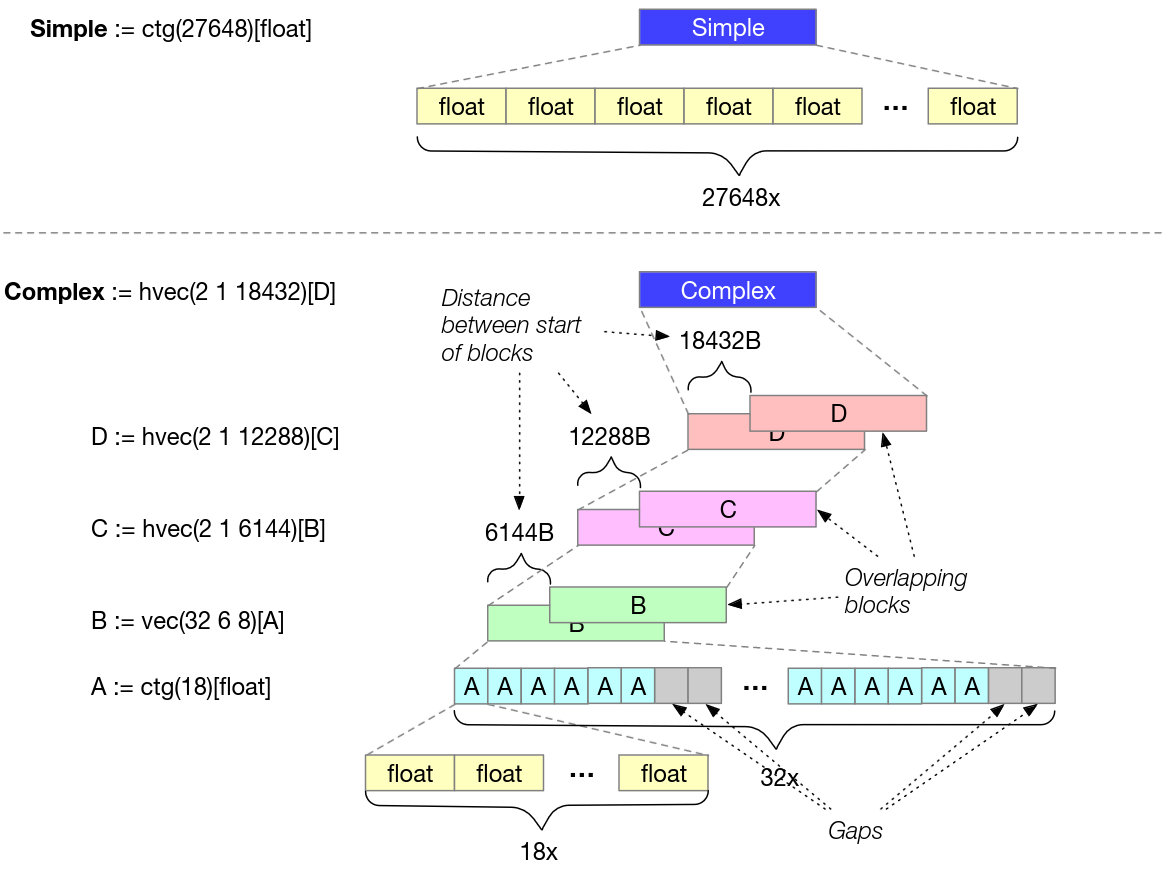}
\caption{The two MPI DDTs used in this demonstration}
\label{fig:mpi_ddt_definition}
\end{figure}

MPI Derived Datatypes~\cite{mpiddt} (DDT) is a mechanism specified by MPI~\cite{mpi} to serialize data from non-contiguous or overlapping buffers into a single MPI message (or de-serialize upon receiving a message). MPI DDTs are built by nesting data-layout specifications.
The basic building blocks are primitive types built into MPI, such as \texttt{MPI\_FLOAT}, which refers to a single 32-bit floating point value. These building blocks are used in conjunction with MPIs DDT constructors, such as \texttt{MPI\_Type\_contiguous(count, oldtype, newtype)} which creates a new DDT  consisting of \texttt{count} consecutive repetitions of \texttt{oldtype}.
MPI specifies many type-constructing functions, in this work we use \texttt{MPI\_Type\_contiguous}, \texttt{MPI\_Type\_vector}, and \texttt{MPI\_Type\_hvector}.
The last two allow the user to specify a count (how many times the pattern repeats), a block length (how many occurrences of \texttt{oldtype} form a block), and a stride (the distance between the starts of two blocks, if the stride is larger than the block length there will be gaps in the de-serialized data on the receiver side, if the stride is smaller, data is repeated multiple times in the message). The difference between the vector and hvector DDT constructor is that in an hvector the stride is given in bytes, while in a vector it is specified as multiples of oldtype. For this demonstration we construct both a simple and a more complex (multiple nested types, overlap between blocks) DDT, both are shown in Figure~\ref{fig:mpi_ddt_definition}.

We vary the message size in the following experiments by sending a varying number of the ``simple'' or ``complex'' DDT per message (increasing the count argument in \texttt{MPI\_Send}).

Previous work~\cite{pspin_ddt} has ported the MPICH dataloop~\cite{dataloop} DDT engine to sPIN handlers. We port these existing sPIN handlers to FPsPIN and
characterise the de-serialization throughput of for different DDTs. Since MPI Datatypes require to send messages of arbitrary length in a reliable fashion, we utilize the SLMP protocol introduced previously. The utilized dataloop engine also relies on in-order processing of the received data. We avoid message-level parallelism by setting the SLMP window-size to one. We re-introduce parallelism by sending multiple (16 in the experiment shown below) MPI messages of the same size and with the same datatype in parallel.

\begin{figure}[!htb]
\centering
\includegraphics[width=\linewidth]{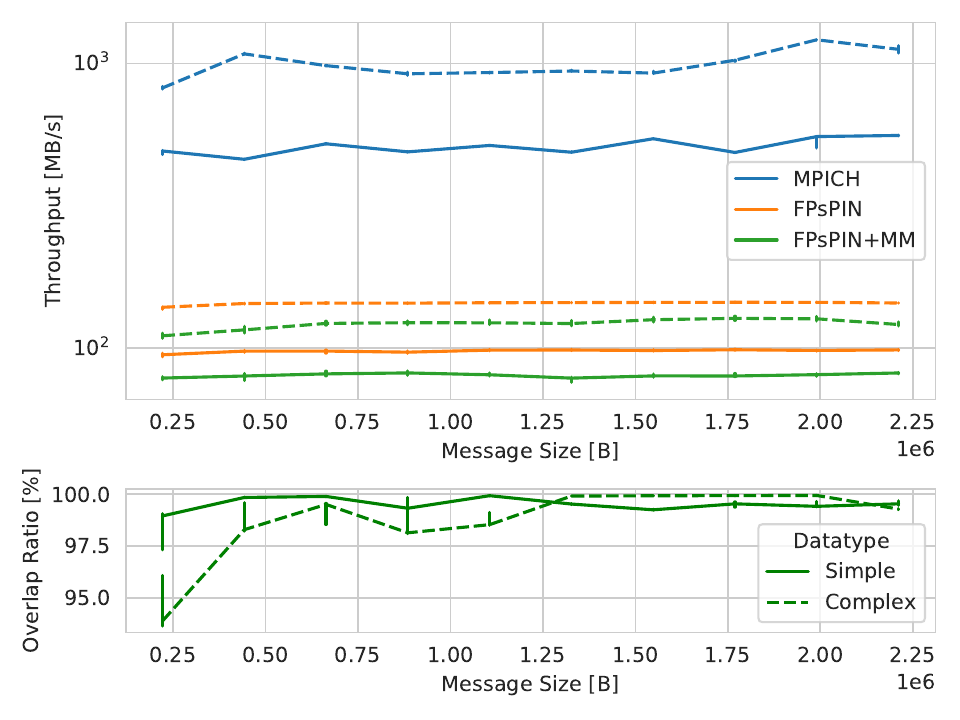}
\caption{Throughput and overlap ratio of MPI DDT processing.}
\label{fig:mpi_ddt_tput_ovlp}
\end{figure}

Offloading MPI DDT processing is desirable because it allows the host CPU to perform computational tasks while data is placed into the host memory by the NIC, even if the data is not contiguous as in most RDMA operations. We demonstrate this by executing a matrix-matrix multiplication on the host CPU while the transfer is in progress. For each experiment we tune the size of the computation so that it last slightly longer than the data transfer. In this experiment we separately measure the time taken
for the computation $T_{MM}$ and the overhead for setting up and polling for completion of the data-transmission $T_{Poll}$. In Figure~\ref{fig:mpi_ddt_tput_ovlp} we show the achieved throughput for the different MPI DDTs, with an without an overlapping computation (the computation in parallel also causes memory traffic) and the achieved overlap ratio $R = T_{MM} / (T_{MM}+T_{Poll})$.

We observe that FPsPIN is not able to match the performance of the MPICH dataloop implementation on the host, for which we see two main reasons: The host CPU runs at a two orders of magnitude higher clock-rate than the FPsPIN HPUs and the dataloop engine has been well tuned for x86 CPUs, while it was simply ported to the RISC-V FPsPIN HPUs as a proof of concept. It has been shown~\cite{ddt_runtime_recomp} that techniques such as runtime code specialization can improve DDT processing performance on x86 CPUs - we would expect similar benefits for FPsPIN. We also see that offloading datatype processing almost completely alleviates the need for the host CPU to perform any work, even for messages of moderate size 98\% overlap is achieved.

\section{Conclusion}

We presented FPsPIN, a smart NIC research platform utilizing the sPIN abstract machine model. We give an overview of the architecture of FPsPIN which should enable users to modify or augment FPsPIN. We presented several use-cases for FPsPIN, from implementing a simple offloaded ping-pong server to offloading MPI datatype processing. The entire research platform is available to the public as open-source software and is very cost-effective to use, all experiments shown within this work have been performed on a single off-the shelf server and a widely commercially available FPGA development board. We believe FPsPIN is capable of enabling the HPC and datacenter community to answer open research questions such as
which parts of the HPC software stack can benefit from NIC offload in practice,
or how can we augment data-center operating systems and sNICs to enable virtualized sNICs and thereby providing multi-tenancy, without breaking quality of service and security guarantees.

Basing FPsPIN on the sPIN abstract machine model ensures that answers to these, and potentially many more, open research questions based on FPsPIN are generally applicable, portable, and not vendor-specific. 

We see two interesting future directions for the FPsPIN platform: the first is to remove the need for specialized hardware, providing a software-only sPIN interface, which will allow application developers to prototype sPIN applications on their laptop. The second is to allow users to choose between different HPU core architectures, while the PsPIN architecture has been shown to be well suited for ASIC implementations of packet processors, it is not ideal for synthesis on an FPGA.





\bibliographystyle{IEEEtran}
\bibliography{IEEEabrv,references}

\begin{thebibliography}{10}
\providecommand{\url}[1]{#1}
\csname url@samestyle\endcsname
\providecommand{\newblock}{\relax}
\providecommand{\bibinfo}[2]{#2}
\providecommand{\BIBentrySTDinterwordspacing}{\spaceskip=0pt\relax}
\providecommand{\BIBentryALTinterwordstretchfactor}{4}
\providecommand{\BIBentryALTinterwordspacing}{\spaceskip=\fontdimen2\font plus
\BIBentryALTinterwordstretchfactor\fontdimen3\font minus
  \fontdimen4\font\relax}
\providecommand{\BIBforeignlanguage}[2]{{%
\expandafter\ifx\csname l@#1\endcsname\relax
\typeout{** WARNING: IEEEtran.bst: No hyphenation pattern has been}%
\typeout{** loaded for the language `#1'. Using the pattern for}%
\typeout{** the default language instead.}%
\else
\language=\csname l@#1\endcsname
\fi
#2}}
\providecommand{\BIBdecl}{\relax}
\BIBdecl

\bibitem{rdma}
A.~Kalia, M.~Kaminsky, and D.~G. Andersen, ``Design guidelines for high
  performance $\{$RDMA$\}$ systems,'' in \emph{2016 USENIX Annual Technical
  Conference (USENIX ATC 16)}, 2016, pp. 437--450.

\bibitem{datacenter_ethernet}
T.~Hoefler, D.~Roweth, K.~Underwood, R.~Alverson, M.~Griswold, V.~Tabatabaee,
  M.~Kalkunte, S.~Anubolu, S.~Shen, M.~McLaren, A.~Kabbani, and S.~Scott,
  ``Data center ethernet and remote direct memory access: Issues at
  hyperscale,'' \emph{Computer}, vol.~56, no.~7, pp. 67--77, 2023.

\bibitem{molka2014main}
D.~Molka, D.~Hackenberg, and R.~Sch{\"o}ne, ``Main memory and cache performance
  of intel sandy bridge and amd bulldozer,'' in \emph{Proceedings of the
  workshop on Memory Systems Performance and Correctness}, 2014, pp. 1--10.

\bibitem{mogul2003tcp}
J.~C. Mogul, ``$\{$TCP$\}$ offload is a dumb idea whose time has come,'' in
  \emph{9th Workshop on Hot Topics in Operating Systems (HotOS IX)}, 2003.

\bibitem{de2021flare}
D.~De~Sensi, S.~Di~Girolamo, S.~Ashkboos, S.~Li, and T.~Hoefler, ``Flare:
  Flexible in-network allreduce,'' in \emph{Proceedings of the International
  Conference for High Performance Computing, Networking, Storage and Analysis},
  2021, pp. 1--16.

\bibitem{istvan2016consensus}
Z.~Istv{\'a}n, D.~Sidler, G.~Alonso, and M.~Vukolic, ``Consensus in a box:
  Inexpensive coordination in hardware,'' in \emph{13th USENIX Symposium on
  Networked Systems Design and Implementation (NSDI 16)}, 2016, pp. 425--438.

\bibitem{spin_orig}
T.~Hoefler, S.~D. Girolamo, K.~Taranov, R.~E. Grant, and R.~Brightwell,
  ``{sPIN: High-performance streaming Processing in the Network},'' in
  \emph{Proceedings of the International Conference for High Performance
  Computing, Networking, Storage and Analysis (SC17)}, Nov. 2017.

\bibitem{pspin}
S.~D. Girolamo, A.~Kurth, A.~Calotoiu, T.~Benz, T.~Schneider, J.~Beránek,
  L.~Benini, and T.~Hoefler, ``{A RISC-V in-network accelerator for flexible
  high-performance low-power packet processing},'' in \emph{Proceedings of the
  48th Annual International Symposium on Computer Architecture (ISCA'21)}, Jun.
  2021.

\bibitem{pspin_protobuf}
S.~Cao, S.~D. Girolamo, and T.~Hoefler, ``{Accelerating Data
  Serialization/Deserialization Protocols with In-Network Compute},'' in
  \emph{2022 IEEE/ACM International Workshop on Exascale MPI (ExaMPI)}, Nov.
  2022.

\bibitem{pspin_fs}
S.~D. Girolamo, D.~D. Sensi, K.~Taranov, M.~Malesevic, M.~Besta, T.~Schneider,
  S.~Kistler, and T.~Hoefler, ``{Building Blocks for Network-Accelerated
  Distributed File Systems},'' in \emph{Proceedings of the International
  Conference for High Performance Computing, Networking, Storage and Analysis
  (SC'22)}, Nov. 2022.

\bibitem{pspin_ddt}
S.~Di~Girolamo, K.~Taranov, A.~Kurth, M.~Schaffner, T.~Schneider,
  J.~Ber{\'a}nek, M.~Besta, L.~Benini, D.~Roweth, and T.~Hoefler,
  ``Network-accelerated non-contiguous memory transfers,'' in \emph{Proceedings
  of the International Conference for High Performance Computing, Networking,
  Storage and Analysis}, 2019, pp. 1--14.

\bibitem{snyder2018verilator}
W.~Snyder, ``Verilator 4.0: open simulation goes multithreaded,'' in \emph{Open
  Source Digital Design Conference (ORConf)}, 2018.

\bibitem{xu_thesis}
P.~Xu, ``\BIBforeignlanguage{en}{Full-system evaluation of the spin
  in-network-compute architecture},'' Master Thesis, ETH Zurich, Zurich,
  2023-09.

\bibitem{wei2023characterizing}
X.~Wei, R.~Cheng, Y.~Yang, R.~Chen, and H.~Chen, ``Characterizing off-path
  $\{$SmartNIC$\}$ for accelerating distributed systems,'' in \emph{17th USENIX
  Symposium on Operating Systems Design and Implementation (OSDI 23)}, 2023,
  pp. 987--1004.

\bibitem{petrini2002quadrics}
F.~Petrini, W.-c. Feng, A.~Hoisie, S.~Coll, and E.~Frachtenberg, ``The quadrics
  network: High-performance clustering technology,'' \emph{Ieee Micro},
  vol.~22, no.~1, pp. 46--57, 2002.

\bibitem{yu2004efficient}
W.~Yu, D.~Buntinas, R.~L. Graham, and D.~K. Panda, ``Efficient and scalable
  barrier over quadrics and myrinet with a new nic-based collective message
  passing protocol,'' in \emph{18th International Parallel and Distributed
  Processing Symposium, 2004. Proceedings.}\hskip 1em plus 0.5em minus
  0.4em\relax IEEE, 2004, p. 182.

\bibitem{wagner2004nic}
A.~Wagner, H.-W. Jin, D.~K. Panda, and R.~Riesen, ``Nic-based offload of
  dynamic user-defined modules for myrinet clusters,'' in \emph{2004 IEEE
  International Conference on Cluster Computing (IEEE Cat. No. 04EX935)}.\hskip
  1em plus 0.5em minus 0.4em\relax IEEE, 2004, pp. 205--214.

\bibitem{bosshart2014p4}
P.~Bosshart, D.~Daly, G.~Gibb, M.~Izzard, N.~McKeown, J.~Rexford,
  C.~Schlesinger, D.~Talayco, A.~Vahdat, G.~Varghese \emph{et~al.}, ``P4:
  Programming protocol-independent packet processors,'' \emph{ACM SIGCOMM
  Computer Communication Review}, vol.~44, no.~3, pp. 87--95, 2014.

\bibitem{catapult}
\BIBentryALTinterwordspacing
D.~Chiou, ``The microsoft catapult project,'' in \emph{2017 IEEE International
  Symposium on Workload Characterization (IISWC)}.\hskip 1em plus 0.5em minus
  0.4em\relax Los Alamitos, CA, USA: IEEE Computer Society, oct 2017, pp.
  124--124. [Online]. Available:
  \url{https://doi.ieeecomputersociety.org/10.1109/IISWC.2017.8167769}
\BIBentrySTDinterwordspacing

\bibitem{forencich2020corundum}
A.~Forencich, A.~C. Snoeren, G.~Porter, and G.~Papen, ``Corundum: An
  open-source 100-gbps nic,'' in \emph{2020 IEEE 28th Annual International
  Symposium on Field-Programmable Custom Computing Machines (FCCM)}.\hskip 1em
  plus 0.5em minus 0.4em\relax IEEE, 2020, pp. 38--46.

\bibitem{stingray}
\BIBentryALTinterwordspacing
Broadcom. Stingray ps225. [Online]. Available:
  \url{\url{https://docs.broadcom.com/doc/PS225-PB}}
\BIBentrySTDinterwordspacing

\bibitem{bluefield}
I.~Burstein, ``Nvidia data center processing unit (dpu) architecture,'' in
  \emph{2021 IEEE Hot Chips 33 Symposium (HCS)}.\hskip 1em plus 0.5em minus
  0.4em\relax IEEE, 2021, pp. 1--20.

\bibitem{xing2022towards}
T.~Xing, H.~Tajbakhsh, I.~Haque, M.~Honda, and A.~Barbalace, ``Towards portable
  end-to-end network performance characterization of smartnics,'' in
  \emph{Proceedings of the 13th ACM SIGOPS Asia-Pacific Workshop on Systems},
  2022, pp. 46--52.

\bibitem{dpdk}
\BIBentryALTinterwordspacing
L.~Foundation, ``Data plane development kit ({DPDK}),'' 2015. [Online].
  Available: \url{http://www.dpdk.org}
\BIBentrySTDinterwordspacing

\bibitem{iso_osi_model}
\BIBentryALTinterwordspacing
J.~Day, ``The (un)revised osi reference model,'' \emph{SIGCOMM Comput. Commun.
  Rev.}, vol.~25, no.~5, p. 39–55, oct 1995. [Online]. Available:
  \url{https://doi.org/10.1145/216701.216704}
\BIBentrySTDinterwordspacing

\bibitem{fpga}
I.~Kuon, R.~Tessier, J.~Rose \emph{et~al.}, ``Fpga architecture: Survey and
  challenges,'' \emph{Foundations and Trends{\textregistered} in Electronic
  Design Automation}, vol.~2, no.~2, pp. 135--253, 2008.

\bibitem{axi1}
ARM, \emph{AMBA 4 AXI4-Stream Protocol}, 2010.

\bibitem{axi2}
A.~Ranga, L.~Venkatesh, and V.~Venkanna, ``Design and implementation of
  amba-axi protocol using vhdl for soc integration,'' \emph{Int J Eng Res
  Appl}, vol.~2, no.~4, pp. 1102--1110, 2012.

\bibitem{waterman2014risc}
A.~Waterman, Y.~Lee, D.~Patterson, K.~Asanovic, V.~I.~U. level Isa,
  A.~Waterman, Y.~Lee, and D.~Patterson, ``The risc-v instruction set manual,''
  \emph{Volume I: User-Level ISA’, version}, vol.~2, pp. 1--79, 2014.

\bibitem{pulp}
F.~Conti, D.~Rossi, A.~Pullini, I.~Loi, and L.~Benini, ``Energy-efficient
  vision on the pulp platform for ultra-low power parallel computing,'' in
  \emph{2014 IEEE Workshop on Signal Processing Systems (SiPS)}.\hskip 1em plus
  0.5em minus 0.4em\relax IEEE, 2014, pp. 1--6.

\bibitem{corundum_doc}
\BIBentryALTinterwordspacing
A.~F. et~al. (2024) Corundum readme. [Online]. Available:
  \url{https://github.com/corundum/corundum}
\BIBentrySTDinterwordspacing

\bibitem{hoffman2003testing}
D.~Hoffman, D.~Prabhakar, and P.~Strooper, ``Testing iptables,'' in
  \emph{Proceedings of the 2003 conference of the Centre for Advanced Studies
  on Collaborative research}, 2003, pp. 80--91.

\bibitem{iptables_u32}
\BIBentryALTinterwordspacing
D.~Cohen and G.~Kessler. (2024) {IPTables} u32 matcher description. [Online].
  Available: \url{http://www.stearns.org/doc/iptables-u32.current.html}
\BIBentrySTDinterwordspacing

\bibitem{john2007analysis}
W.~John and S.~Tafvelin, ``Analysis of internet backbone traffic and header
  anomalies observed,'' in \emph{Proceedings of the 7th ACM SIGCOMM conference
  on Internet measurement}, 2007, pp. 111--116.

\bibitem{benson2010understanding}
T.~Benson, A.~Anand, A.~Akella, and M.~Zhang, ``Understanding data center
  traffic characteristics,'' \emph{ACM SIGCOMM Computer Communication Review},
  vol.~40, no.~1, pp. 92--99, 2010.

\bibitem{pulp_sdk}
\BIBentryALTinterwordspacing
N.~Bruschi, G.~Haugou, G.~Tagliavini, F.~Conti, L.~Benini, and D.~Rossi. Pulp
  sdk. [Online]. Available:
  \url{\url{https://github.com/pulp-platform/pulp-sdk}}
\BIBentrySTDinterwordspacing

\bibitem{nazarewicz2019deep}
\BIBentryALTinterwordspacing
M.~Nazarewicz, ``A deep dive into cma,'' \emph{LWN.net}, 2019. [Online].
  Available: \url{https://lwn.net/Articles/486301/}
\BIBentrySTDinterwordspacing

\bibitem{infiniband}
G.~F. Pfister, ``An introduction to the infiniband architecture,'' \emph{High
  performance mass storage and parallel I/O}, vol.~42, no. 617-632, p.~10,
  2001.

\bibitem{roce}
G.~Kaur and M.~Bala, ``Rdma over converged ethernet: A review,''
  \emph{International Journal of Advances in Engineering \& Technology},
  vol.~6, no.~4, p. 1890, 2013.

\bibitem{iperf}
A.~Tirumala, ``Iperf: The tcp/udp bandwidth measurement tool,''
  \emph{http://dast. nlanr. net/Projects/Iperf/}, 1999.

\bibitem{mpiddt}
Q.~Xiong, P.~V. Bangalore, A.~Skjellum, and M.~Herbordt, ``Mpi derived
  datatypes: Performance and portability issues,'' in \emph{Proceedings of the
  25th European MPI Users' Group Meeting}, 2018, pp. 1--10.

\bibitem{mpi}
\BIBentryALTinterwordspacing
{Message Passing Interface Forum}, \emph{{MPI}: A Message-Passing Interface
  Standard Version 4.0}, Jun. 2021. [Online]. Available:
  \url{https://www.mpi-forum.org/docs/mpi-4.0/mpi40-report.pdf}
\BIBentrySTDinterwordspacing

\bibitem{dataloop}
R.~Ross, N.~Miller, and W.~D. Gropp, ``Implementing fast and reusable datatype
  processing,'' in \emph{European Parallel Virtual Machine/Message Passing
  Interface Users’ Group Meeting}.\hskip 1em plus 0.5em minus 0.4em\relax
  Springer, 2003, pp. 404--413.

\bibitem{ddt_runtime_recomp}
T.~Schneider, F.~Kjolstad, and T.~Hoefler, ``Mpi datatype processing using
  runtime compilation,'' in \emph{Proceedings of the 20th European MPI Users'
  Group Meeting}, 2013, pp. 19--24.

\end{thebibliography}

\end{document}